\documentclass[journal]{IEEEtran}

\ifCLASSINFOpdf
\else
   \usepackage[dvips]{graphicx}
\fi
\usepackage{url}

\usepackage{supertabular}
\usepackage{kbordermatrix}
\usepackage{graphicx}
\usepackage{amsmath,empheq}
\usepackage{algorithm}
\usepackage{algorithmicx}
\usepackage{algpseudocode}
\usepackage{booktabs}
\usepackage{longtable}
\usepackage{mathtools}
\usepackage{amssymb}
\usepackage[table]{xcolor}
\mathtoolsset{showonlyrefs} 
\definecolor{light-gray}{HTML}{FFFFFF}
\definecolor{light-cyan}{HTML}{C4C4C4}

\newcommand{\minuseq}{\mathrel{-}=}
\newcommand{\producteq}{\mathrel{*}=}

\newcommand\x[1]{%
  \colorbox{gray!30}{$#1$}%
}

\newcommand\y[1]{%
  \colorbox{gray!70}{$#1$}%
}

\newcommand\z[1]{%
  \colorbox{black!70}{$#1$}%
}

\begin{document}
\title{On the Generation of Long Binary Sequences with Record-Breaking PSL Values}

\author{Miroslav Dimitrov \IEEEmembership{Student Member, IEEE}, Tsonka Baitcheva \IEEEmembership{Member, IEEE}, and Nikolay Nikolov
\thanks{This work has been partially supported by the Bulgarian National Science Fund under contract number DH 12/8, 15.12.2017.}
\thanks{M. Dimitrov  and T.  Baicheva are with the Institute of Mathematics and Informatics, Bulgarian Academy of Sciences, Sofia, Bulgaria (email:mirdim@math.bas.bg, email:tsonka@math.bas.bg).}
\thanks{N. Nikolov is with the State Agency of National Security, Sofia, Bulgaria.}}%

\maketitle

\begin{abstract}
Binary sequences are widely used in various practical fields, such as telecommunications, radar technology, navigation, cryptography, measurement sciences, biology or industry. In this paper, a  method to generate long binary sequences (LBS) with low peak sidelobe level (PSL) value is proposed. Having an LBS with length $n$, both the time and memory complexities of the proposed algorithm are $\mathcal{O}(n)$. During our experiments, we repeatedly reach better PSL values than the currently known state of art constructions, such as Legendre sequences, with or without rotations, Rudin-Shapiro sequences or m-sequences, with or without rotations, by always reaching a record-breaking PSL values strictly less than $\sqrt{n}$. Furthermore, the efficiency and simplicity of the proposed method are particularly beneficial to the lightweightness of the implementation, which allowed us to reach record-breaking PSL values for less than a second.
\end{abstract}


\IEEEpeerreviewmaketitle

\section{Introduction}

\IEEEPARstart{T}{he} practical fields in which binary sequences with low PSL values could be exploitable are manifold \cite{golomb2005signal}. Despite the widely usage of binary sequences in radar and sonar pulse compression systems \cite{kroszczynski1969pulse}, they are further used, for example, in the estimation of the shape of hemodynamic responses \cite{buracas2002efficient}, analysis of visual neurons \cite{reid1997use}, audio watermarking \cite{cvejic2001audio}, orthogonal frequency division multiplexing \cite{tellambura1997use}, CDMA systems \cite{ulukus1998optimum}, scrambling algorithms \cite{zhou2008image}, motion tracking technologies \cite{xiao2013lumitrack} and many others.

Of special interest are those binary sequences possessing low PSL value. Some strategies for construction of such sequences comprise the Barker codes \cite{barker1953group}, Rudin-Shapiro sequences \cite{rudin1959some}\cite{shapiro1952extremal}, maximal length shift register sequences, or m-sequences  \cite{golomb1967shift}, Gold codes \cite{gold1967optimal}, Kasami codes \cite{kasami1966weight}, Weil sequences \cite{rushanan2006weil}, Legendre sets \cite{pott2006finite} and others (see \cite{levanon2004radar}). 

M-sequences, Gold codes and Kasami sequences have ideal periodic autocorrelation functions but have no constraints on the sidelobes of their aperiodic autocorrelation functions, i.e. their PSL value is not pre-determined. Same is true for Legendre sets and Rudin-Shapiro sequences. Furthermore, it is difficult to calculate the growth of the PSL of the aforementioned families of binary sequences. It is conjectured that the PSL values of $m$-sequences grow like $\mathcal{O}(\sqrt{n})$, making them one of the best methods to straightforwardly construct a binary sequences with near-optimal PSL value. However, as stated in \cite{jedwab2006peak}:

\begin{quote}
\textit{The claim that the PSL of $m$-sequences grows like $\mathcal{O}(\sqrt{n})$, which appears frequently in the radar literature, is concluded to be unproven and not currently supported by data.}
\end{quote}

As summarized in \cite{nasrabadi2010survey}, during the years a variety of analytical constructions and computer search methods are developed in order to construct binary sequences with relatively small or minimal PSL. By an exhaustive search the minimum values of the PSL for $n\leq 40$\cite{lindner1975binary}, $n\leq 48$\cite{baden1990optimal}, $n=64$\cite{coxson2005efficient}, $n\leq 68$\cite{leukhin2012binary}, $n\leq 74$\cite{leukhin2013optimal},  $n\leq 80$ \cite{leukhin2014exhaustive}, $n\leq 82$ \cite{leukhin2015bernasconi} and $n\leq 84$ \cite{leukhin2017exhaustive} are obtained. The best currently known values for PSL for $ 85 \leq n \leq 105$ are published in \cite{NC}, and for some discrete values of $n \geq 106$ in \cite{dzvonkovskaya2008long}\cite{Patent}\cite{mow2015new}\cite{dimitrov2020efficient}. 

It appears that the current state of art computer search methods, like CAN \cite{he2009designing}, ITROX \cite{soltanalian2012computational}, MWISL-Diag, MM-PSL \cite{song2015sequence} or DPM \cite{kerahroodi2017coordinate}, could yield better, or at least not worse PSL values, than the algebraic constructions. However, when the length of the generated by a given heuristic algorithm binary sequences rises, so is the overall time and memory complexity of the routine. As concluded in \cite{mow2015new}:

\begin{quote}
\textit{As an indication of the runtime complexity of our EA\footnote{EA stands for Evolutionary Algorithm}, the computing time is 58009 s or 16.1136 h for L=1019. For lengths up to 4096, the computing time required empirically shows a seemingly quadratic growth with L.}
\end{quote}

Thus, the main motivation of this work is to create an efficient and lightweight algorithm, in terms of time and memory complexity, to address the heuristic generation of very long binary sequences with near-optimal PSL values.

\section{Preliminaries}
\label{sec:prelims}

Let $B=(b_0,b_1,\cdots ,b_{n-1})$ be a binary sequence of length $n>1$, where $b_i\in \{-1,1\}, 0\leq i\leq n-1$. The aperiodic autocorrelation function of $B$, or \textbf{AACF}, is given by $$C_u(B)=\sum_{j=0}^{n-u-1} b_jb_{j+u}, \ \ for \ u\in \{0,1,\cdots, n-1\}.$$  

$C_0(B)$ is called \textbf{mainlobe} and the rest $C_u(B)$ for $u\in \{1, \cdots ,n-1\}$ are called \textbf{sidelobe} levels. We define the peak sidelobe level of $B$, or \textbf{PSL} \cite{turyn1968sequences}, as $$B_{PSL}=\max_{0<u<n} \lvert C_u(B)\rvert.$$

\section{Some observations about the PSL calculation}
\label{sec:observationsPartI}

Let us denote $C_{n-i-1}(B)$ by $\hat{C}_i(B)$. Since this is just a rearrangement of the sidelobes of $B$, it follows that:
\[
	B_{PSL}=\max_{0<u<n} \lvert C_u(B)\rvert = \max_{0 \le u < n-1} \lvert \hat{C}_u(B)\rvert.
\]

\begin{figure}[h]
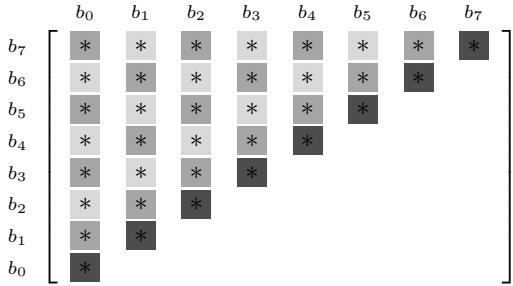

\renewcommand{\kbldelim}{[}
\renewcommand{\kbrdelim}{]}

\[
   \kbordermatrix{
    & b_0 & b_1 & b_2 & b_3 & b_4 & b_5 & b_6 & b_7\\
    b_7 & \y * &\x * & \y * & \x * &\y * &\x * &\y * &\z * \\
    b_6 &  \x * & \y * & \x * &\y * &\x * &\y * &\z * \\
    b_5 & \y * &\x * & \y * & \x * &\y * &\z * \\
    b_4 &  \x * & \y * & \x * &\y * &\z * \\
    b_3 & \y * &\x * & \y * & \z * \\
    b_2 & \x * & \y * & \z * \\
    b_1 & \y * &\z *\\
    b_0 & \z * \\
  }
\]
\caption{A visual interpretation of the sidelobe calculation process, for a binary sequence with length 8}
\label{fig:SideLobes8}
\end{figure}

We will graphically represent the calculation of values of $\hat{C}_i(B)$ for a binary sequence of length 8 in Figure \ref{fig:SideLobes8}. The $x$-axis indexes represent the elements of $B =(b_0, b_1, \cdots, b_7)$, while the $y$ axes represents the elements of $B$ in reverse order, i.e. $(b_7, b_6, \cdots, b_0)$. Each cell of the graphics corresponds to the product of the provided by the $x$ and $y$-axis values. To calculate $\hat{C}_i(B)=\sum_{j=0}^{i} b_jb_{j+n-i-1}$ for some $i$ ($0 \leq i \leq 7$), we start from the cell with coordinates $(b_i,b_7)$. Then, by decreasing both indexes of the current cell by $1$ we jump to the next cell $(b_{i-1},b_6)$ which will be added to the sum. We continue this process until we reach the cell $(b_0,b_{7-i})$.

As the value of the mainlobe $\hat{C}_7(B)$ is always 8, we can exclude it from the PSL calculation. Having this in mind, we can define the PSL of the binary sequence $B$ as the diagonal in Figure \ref{fig:SideLobes8} with the highest absolute sum of its elements compared to all other diagonals, excluding the main one.

Let us denote by $\overline{b_i}$ the flipped bit $b_i$, i.e.  $\overline{b_i} = -b_i$ and by $\hat{C}_i(B_j)$ the sidelobe of the binary sequence $B_j$, obtained from $B$ by flipping the bit on position $j$. 

We can further exploit the relations between the value of the sidelobe $\hat{C}_i(\Psi)$ of a given binary sequence $\Psi$ with length $n$, and the value of the sidelobe $\hat{C}_i(\Psi_f)$, s.t. the binary sequence $\Psi_f$ is equal to the binary sequence $\Psi$ with the bit on position $f$ flipped. We denote as $\Omega_{\Psi}$ the array of all the consequent sidelobes of $\Psi$, i.e: 
\[ 
\Omega_{\Psi} = \left[ \hat{C}_0(\Psi), \hat{C}_1(\Psi), \cdots, \hat{C}_{n-2}(\Psi) \right]
\]

We denote as $\Omega_{\Psi_{f}}$ the array of all the consequent sidelobes of $\Psi_{f}$, i.e:
\[ 
\Omega_{\Psi_f} = \left[ \hat{C}_0(\Psi_f), \hat{C}_1(\Psi_f), \cdots, \hat{C}_{n-2}(\Psi_f) \right]
\]

For convenience, we further denote the $i$-th element of a given array $A$ as $A[i]$. For example, $\Omega_{\Psi}[3] = \hat{C}_2(\Psi)$. 

The calculation of $\Omega_{\Psi}$, corresponding to some random binary sequence $\Psi$, is not linear. The time complexity of the trivial computational approach is $\mathcal{O}(n^2)$ (two nested \textbf{for} cycles). However, as shown in Wiener--Khinchin--Einstein theorem \cite{wiener1964extrapolation}, the autocorrelation function of a wide-sense-stationary random process has a spectral decomposition given by the power spectrum of that process, we can use one regular and one inverse Fast Fourier Transform (FFT), to achieve a faster way of calculating $\Omega_{\Psi}$. Despite its time complexity of $\mathcal{O}(n\log{n})$, its memory complexity is significantly higher than the trivial computational approach.

By exploiting the observations made in this section, we present an algorithm which can calculate the array $\Omega_{\Psi_f}$, if we hold the array $\Omega_{\Psi}$ in memory, with time and memory complexity of $\mathcal{O}(n)$.  The pseudo-code of the algorithm is given in Algorithm \ref{algor:InMemoryFlip}. The following notations are used:

\begin{itemize}
\item{$\min_{\left(x, y\right)}$ : returns $x$, if $x \leq y$; otherwise, returns $y$.}
\item{$\max_{\left(x, y\right)}$ : returns $x$, if $x \geq y$; otherwise, returns $y$.}
\item{$x \minuseq y$ : same as $x=x-y$}
\item{$x \producteq y$ : same as $x=x*y$}
\end{itemize}

\algrenewcommand\algorithmicindent{0.5em}%
\begin{algorithm}[]
\caption{In-memory flip}
\label{algor:InMemoryFlip}
\begin{algorithmic}[1]
\Procedure{Flip}{$f, \Psi, \Omega_{\Psi}, n$}
\State $\delta_{min} \gets \min_{\left(n-f-1, f\right)}$
\State $\delta_{max} \gets \max_{\left(n-f, f \right)}$
\If{$f \leq \frac{n-1}{2}$}
	\For{$q \in \left[0, \delta_{max}-\delta_{min}-1\right)$}
	\State $\Omega_{\Psi}[\delta_{min}+q] \minuseq 2\Psi[f]\Psi[n-q-1]$
	\EndFor
\Else
	\For{$q \in \left[0, \delta_{max}-\delta_{min}\right)$}
	\State $\Omega_{\Psi}[\delta_{min}+q] \minuseq 2\Psi[f]\Psi[q]$
	\EndFor	
\EndIf

\If{$f \leq \frac{n-1}{2}$}
	\For{$q \in \left[0, n-\delta_{max}\right)$}
	\State $\Omega_{\Psi}[\delta_{max}+q-1] \minuseq 2\Psi[f]\left(\Psi[2f-q]+\Psi[q]\right)$
	\EndFor
\Else
	\For{$q \in \left[0, n-\delta_{max}-1\right)$}
		\State {$\Omega_{\Psi}[\delta_{max}+q] \minuseq$} \State {$2\Psi[f]\left(\Psi[\delta_{max}-\delta_{min}+q]+\Psi[n-q-1]\right)$}
	\EndFor
\EndIf
\State $\Psi[f] \producteq -1 $
\EndProcedure
\end{algorithmic}
\end{algorithm}  

The procedure introduced in Algorithm \ref{algor:InMemoryFlip} performs an in-place memory update of $\Omega_{\Psi}$, when a single bit on position $f$ of $\Psi$ is flipped. Therefore, when the procedure ends, both $\Psi$ and $\Omega$ are  transformed to $\Psi_f$ and $\Omega_{\Psi_{f}}$. We will note that the procedure is reversible, i.e. if an in-place memory update of $\Omega_{\Psi_{f}}$ is made, when a single bit on position $f$ of $\Psi_f$ is flipped, both $\Psi_f$ and $\Omega_{\Psi_{f}}$ are transformed back to $\Psi$ and $\Omega_{\Psi}$.

\section{Algorithm for finding very long binary sequences with low PSL values}
\label{sec:observationsPartII}

The basic ingredients of some heuristic algorithm could be summarized as:

\begin{itemize}
\item{$\mathcal{A}$: \textbf{metaheuristic algorithm}, like hill climbing, simulated annealing, tabu search, etc.}
\item{$\mathcal{I}$: \textbf{search operator}, which is used to generate the candidates}
\item{$\mathcal{F}$: \textbf{fitness function}, which is used to compare the candidates}
\end{itemize}

In our previous work \cite{dimitrov2020efficient}, we have used shotgun hill climbing as $\mathcal{A}$, a neighborhood search as $\mathcal{I}$, and the following fitness function as $\mathcal{F}$:
\[ F(B) = \sum_{u=1}^{n-1}{\mathopen| C_u(B) \mathclose| ^4} = \sum_{u=1}^{n-1}{\left(\mathopen| {\sum_{j=0}^{n-u-1} b_jb_{j+u}}\mathclose|\right)^4},\]

where $B$ is a binary sequence with length $n$. However, using shotgun hill climbing metaheuristic algorithm for finding very long binary sequences with low PSL is not time efficient because the number of hops required to reach some local optimum, grows exponentially when the length of the binary sequence increases. 

Using a neighborhood search to consequently pick the best candidate among all neighbors could be beneficial in finding LBS with low PSL. However, in the aspect of very long binary sequences this search strategy is extremely slow. For example, in the case of a binary sequence with length $2^{16}$, and $\mathcal{I}$ equivalent to a single flip, in each optimization step we need to fitness all the $2^{16}$ neighbors of the current state $S$ and to pick the one with the best score yielded by $\mathcal{F}$. This observation is still true, even if all the neighbors of $S$ have better scores.

To overcome the disadvantages mentioned above, we choose the following strategy:
\begin{itemize}
\item{$\mathcal{A}$: stochastic hill climbing metaheuristic algorithm. We visit a random neighbor of the current state $S$ and accept it, if it is a better candidate than $S$. Otherwise, we pick another neighbor of $S$ and repeat the process.}
\item{$\mathcal{I}$: we choose a single flip as the search operator, so we can exploit memory and time efficiency of Algorithm \ref{algor:InMemoryFlip}.}
\item{$\mathcal{F}$: since $\hat{C}(B)$s are rearrangements of the sidelobes of $B$, we can use the same fitness function $F(B)$ as in \cite{dimitrov2020efficient}, i.e: \[ F(B) = \sum_{u=0}^{n-2}{\mathopen| \hat{C}_u(B) \mathclose| ^4 = \sum_{u=0}^{n-2}{\hat{C}_u(B)}^4} \] }
\end{itemize}

We need to further address the strategy described in $\mathcal{A}$ of picking the next candidate, or neighbor, of $S$. Let us consider an approach of consistently probing $x$ pseudo-randomly chosen neighbors. In case a better candidate is found, we accept it; otherwise, we try again, until we have accumulated a total number of $t$ consequent fails. Then, we announce that we have reached a local optimum. This model can be described by the Bernoulli distribution. The probability to achieve exactly $r$ successes in $N$ trials is equal to:
\[
P\left(X=r\right) = {N \choose r}p^rq^{N-r},
\]
where $p$ and $q$ are the probabilities of success and failure respectively, i.e. $q=1-p$. We can easily calculate $P\left(X=0\right)$:
\[
P\left(X=0\right) = {N \choose 0}p^0q^{N-0} = q^N = (1-p)^N
\]
We further calculate $P\left(X \geq 1\right)$:

\[
P\left(X \geq 1\right) = 1 - P\left(X=0\right) = 1 - (1-p)^N
\]

Thus, relaying solely on pseudorandom choices of  neighbors is not efficient and there is always a chance to miss the better candidate. We can increase the probability of finding the eventual better candidate, but that significantly overhead the optimization process. Missing a better candidate is undesirable behavior of the optimization  process, specially when we are dealing with very long binary sequences. 

The number of neighbors of a binary sequence $B$ with length $n$ is $n$. Let us denote those neighbors as $i_1, i_2, \cdots, i_n$, where the $j$-th neighbor $i_j$ is equal to $B$ with flipped bit on position $j$. We suggest the following simple search strategy::

\begin{enumerate}
\item{we pick a pseudorandomly generated neighbor $i_r$}
\item{we consequently try, for all $x \in \left[1,n-1\right]$, the neighbors $i_{\left(r+x\right) \mod n }$}
\end{enumerate}

We want to emphasize on the extreme situation when the local optimum is already reached, i.e. $k=0$. The suggested search strategy will detect that in exactly $n$ steps, which is an optimal scenario. Furthermore and more importantly, we never miss a better candidate, if any, and  we keep the non-deterministic nature of the search routine at the same time. 

We suggest Algorithm \ref{algor:VLBSsearch} for finding very long binary sequences with low PSL which is based on the above described $\left( \mathcal{A}, \mathcal{I}, \mathcal{F} \right)$.  The following notations and functions are used in the pseudo-code:
\begin{itemize}
\item{$\Psi$ is a random (initial) binary sequence.}
\item{$x,y \gets a,b$ is equivalent to the statements $x=a$ and $y=b$.}
\item{$R\left(n\right)$ : a function, which generates a pseudo-random integer number $\in \left[0, n\right)$.}
\item{$Q\left(x, B, \Omega_{B}\right)$ : a function, which makes $x$ flips at random bit positions in $B$. We pass $\Omega_{B}$ as argument, so we can use the in-place memory function $Flip$. We apply this function to escape the local minimum, when we are stuck in such.}
\item{\textbf{beacon} : we further implant a beacon in the cost function $F$, so we can simultaneously calculate the PSL of the given binary sequence. Such approach adds a negligible overhead, if any, to the cost function routine.}
\end{itemize}

\algrenewcommand\algorithmicindent{0.5em}%
\begin{algorithm}[]
\caption{An algorithm for binary sequences PSL optimization}
\label{algor:VLBSsearch}
\begin{algorithmic}[1]

\State BestCost, Cost $\gets $ $F(\Omega_{\Psi})$, 0
\State isGImpr, isLImpr $\gets$ True, False
\While {true}
	\If{isGImpr}
		\State r $\gets$ R(n)
		\For{$i \in \left[0, n\right)$}
			\State $Flip\left(\left(r+i\right)\%n, \Psi, \Omega_{\Psi}, n\right)$
			\State Cost $\gets F(\Omega_{\Psi})$ \Comment{ * the beacon is here *}
			\If{BestCost $>$ Cost}
				\State BestCost $\gets$ Cost
				\State isLImpr $\gets$ True
				\State \textbf{break}
			\Else
				\State $Flip\left(\left(r+i\right)\%n, \Psi, \Omega_{\Psi}, n\right)$
			\EndIf
		\EndFor
		\If {isLImpr}
			\State isGImpr, isLImpr $\gets$ True, False
			\State \textbf{continue}
		\Else
			\State isGImpr $\gets$ False
		\EndIf
	\Else
		\State r $\gets$ R(4)
		\State Q(1+r, $\Psi$, $\Omega_{\Psi}$)
		\State isGImpr, isLImpr $\gets$ True, False
	\EndIf
\EndWhile
\end{algorithmic}
\end{algorithm}  

\section{On the complexity of Algorithm \ref{algor:VLBSsearch}}
\label{sec:algComplexity}

We emphasize that the complexity of Algorithm \ref{algor:VLBSsearch} mainly depends on the complexity of  
Algorithm \ref{algor:InMemoryFlip}, because in each iteration during the optimization process, Algorithm \ref{algor:InMemoryFlip} is called twice, in case the new candidate is worse than the current one, and once, if the new candidate is better. The in-memory flip function applied in Algorithm \ref{algor:InMemoryFlip} passes only once through $\Omega_{\Psi}$ array, without creating any memory overheads, to reach time and memory complexities of $\mathcal{O}(n)$. The same observation is true for the simple cost function $F$ - it passes only once through $\Omega_{\Psi}$ to sum all quadrupled values of its elements. The function $Q$ is a random number of calls of $F$ (between 1 and 4). The remaining part of Algorithm \ref{algor:VLBSsearch} consists of a simple automata, which rules the continuous optimization process. Therefore, both time and memory complexities of Algorithm \ref{algor:VLBSsearch} are $\mathcal{O}(n)$.

\section{Results}
\label{sec:results}

We have implemented Algorithm \ref{algor:VLBSsearch} by using the C language and a mid-range computer station. Given the linear time and memory complexity of the algorithm, we were able to repeatedly generate binary sequences with record-breaking PSL values for less than a second. As stated in \cite{mow2015new}, the time required to reach a PSL value 26, for a binary sequence with length 1019, is 58009 seconds or 16.1136 hours. For comparison, by using Algorithm \ref{algor:VLBSsearch}, we reach this value for less than a second. 

\begin{figure}[h]
\includegraphics[width=0.5\textwidth]{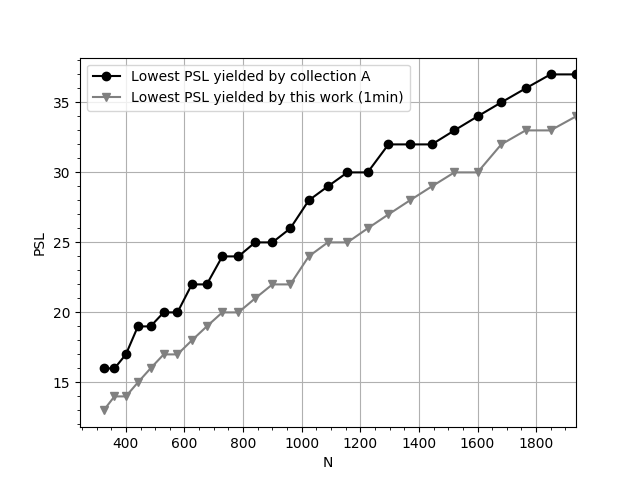}
\caption{Comparison to other state of the art algorithms known in literature}
\label{fig:Comp}
\end{figure}

We present the results achieved by Algorithm \ref{algor:VLBSsearch}, for binary sequences with lengths $x^2$ for $x \in \left[ 18, 44\right]$, compared with the currently known state of art algorithms found in the literature, like CAN \cite{he2009designing}, ITROX \cite{soltanalian2012computational}, MWISL-Diag, MM-PSL \cite{song2015sequence}, DPM \cite{kerahroodi2017coordinate}, 1bCAN \cite{lin2019efficient}. We will refer to this collection of algorithms as collection \textbf{A}. We want to emphasize, that the differences between the proposed algorithm with algorithms from collection A are manifold. For example, we do not use converging functions, mini regular or quadratic optimization problems, floating-based arithmetic. Furthermore, the provided algorithm does not suffer from an unique navigation trace through the sequence search space. The experiments were based on 12 instances of each algorithm (each ran to a distinct thread of the processor). Furthermore, the lifetime of our algorithm is restricted to 1 minute. As shown in Figure \ref{fig:Comp}, we significantly outperform the best results achieved by state of art algorithms. In fact, for some of the lengths, less than a second was needed to reach a record-breaking PSL.
 
In contrast to some other state of the art algorithms, the computing complexity of the algorithm presented in this work does not grow quadratically. Maybe this is the reason for the lack of published results for binary sequences of lengths greater than $2^{12}$. Nevertheless, the results with which we can further compare are $m$-sequences. However, such sequences exists only for lengths $2^n-1$, $n \geq 1, n \in N$.

In Table \ref{tab:mSeqComparisonTimes} we present the best PSL values of binary m-sequences with length $n$ (with or without rotation), yielded by some primitive polynomial of degree $n$ over $GF(2)$ from \cite{dmitriev2007bounds} denoted by $\mathbb{M}^\mathbb{F}_n$ and the  binary sequences generated by Algorithm \ref{algor:VLBSsearch} denoted by $\mathbb{A}_n$ for lengths $2^n-1$ and $13 \leq n \leq 17$. As it can be seen from Table \ref{tab:mSeqComparisonTimes}, our results significantly outperform the best results achieved by m-sequences. 

\begin{table}
\begin{center}
\caption{Reached PSL values compared to known results from m-sequences exhaustive search}
\label{tab:mSeqComparisonTimes}
\ttfamily
\rowcolors{2}{light-gray}{light-cyan}
\begin{tabular}{lcccl}
$n$ & $2^n-1$ &  $\mathbb{M}^\mathbb{F}_n$ &  $\mathbb{A}_n$ \\
\toprule
13 & 8191 & 85 & 77 \\
14 & 16383 & 125 & 115\\
15 & 32767 & 175 & 171\\
16 & 65535 & 258 & 254 \\
17 & 131071 & 363 & 360\\
\midrule
\showrowcolors
\bottomrule
\end{tabular}
\end{center}
\end{table}

\section{Conclusions}
In this paper we present an efficient heuristic algorithm for finding very long binary sequences with record-breaking PSL values. Since the time and memory complexities of the suggested algorithm are both $\mathcal{O}(n)$,  we were able to construct binary sequences with record-breaking PSL values for less than a second. 
\clearpage

\vfill

\bibliographystyle{IEEEtran}
\bibliography{refs}

\begin{thebibliography}{10}
\providecommand{\url}[1]{#1}
\csname url@samestyle\endcsname
\providecommand{\newblock}{\relax}
\providecommand{\bibinfo}[2]{#2}
\providecommand{\BIBentrySTDinterwordspacing}{\spaceskip=0pt\relax}
\providecommand{\BIBentryALTinterwordstretchfactor}{4}
\providecommand{\BIBentryALTinterwordspacing}{\spaceskip=\fontdimen2\font plus
\BIBentryALTinterwordstretchfactor\fontdimen3\font minus
  \fontdimen4\font\relax}
\providecommand{\BIBforeignlanguage}[2]{{%
\expandafter\ifx\csname l@#1\endcsname\relax
\typeout{** WARNING: IEEEtran.bst: No hyphenation pattern has been}%
\typeout{** loaded for the language `#1'. Using the pattern for}%
\typeout{** the default language instead.}%
\else
\language=\csname l@#1\endcsname
\fi
#2}}
\providecommand{\BIBdecl}{\relax}
\BIBdecl

\bibitem{golomb2005signal}
S.~W. Golomb and G.~Gong, \emph{Signal design for good correlation: for
  wireless communication, cryptography, and radar}.\hskip 1em plus 0.5em minus
  0.4em\relax Cambridge University Press, 2005.

\bibitem{kroszczynski1969pulse}
J.~J. Kroszczynski, ``Pulse compression by means of linear-period modulation,''
  \emph{Proceedings of the IEEE}, vol.~57, no.~7, pp. 1260--1266, 1969.

\bibitem{buracas2002efficient}
G.~T. Buracas and G.~M. Boynton, ``Efficient design of event-related fmri
  experiments using m-sequences,'' \emph{Neuroimage}, vol.~16, no.~3, pp.
  801--813, 2002.

\bibitem{reid1997use}
R.~C. Reid, J.~Victor, and R.~Shapley, ``The use of m-sequences in the analysis
  of visual neurons: linear receptive field properties,'' \emph{Visual
  neuroscience}, vol.~14, no.~6, pp. 1015--1027, 1997.

\bibitem{cvejic2001audio}
N.~Cvejic, A.~Keskinarkaus, and T.~Seppanen, ``Audio watermarking using
  m-sequences and temporal masking,'' in \emph{Proceedings of the 2001 IEEE
  Workshop on the Applications of Signal Processing to Audio and Acoustics
  (Cat. No. 01TH8575)}.\hskip 1em plus 0.5em minus 0.4em\relax IEEE, 2001, pp.
  227--230.

\bibitem{tellambura1997use}
C.~Tellambura, ``Use of m-sequences for ofdm peak-to-average power ratio
  reduction,'' \emph{Electronics Letters}, vol.~33, no.~15, pp. 1300--1301,
  1997.

\bibitem{ulukus1998optimum}
S.~Ulukus and R.~D. Yates, ``Optimum multiuser detection is tractable for
  synchronous cdma systems using m-sequences,'' \emph{IEEE Communications
  Letters}, vol.~2, no.~4, pp. 89--91, 1998.

\bibitem{zhou2008image}
Y.~Zhou, K.~Panetta, and S.~Agaian, ``An image scrambling algorithm using
  parameter bases m-sequences,'' in \emph{2008 International Conference on
  Machine Learning and Cybernetics}, vol.~7.\hskip 1em plus 0.5em minus
  0.4em\relax IEEE, 2008, pp. 3695--3698.

\bibitem{xiao2013lumitrack}
R.~Xiao, C.~Harrison, K.~D. Willis, I.~Poupyrev, and S.~E. Hudson, ``Lumitrack:
  low cost, high precision, high speed tracking with projected m-sequences,''
  in \emph{Proceedings of the 26th annual ACM symposium on User interface
  software and technology}, 2013, pp. 3--12.

\bibitem{barker1953group}
R.~H. Barker and W.~Jackson, ``{Group synchronization of binary digital systems
  in Communication Theory},'' \emph{Academic Press, New York}, pp. 273--287,
  1953.

\bibitem{rudin1959some}
W.~Rudin, ``Some theorems on fourier coefficients,'' \emph{Proceedings of the
  American Mathematical Society}, vol.~10, no.~6, pp. 855--859, 1959.

\bibitem{shapiro1952extremal}
H.~S. Shapiro, ``Extremal problems for polynomials and power series,'' Ph.D.
  dissertation, Massachusetts Institute of Technology, 1952.

\bibitem{golomb1967shift}
S.~W. Golomb \emph{et~al.}, \emph{{Shift register sequences}}.\hskip 1em plus
  0.5em minus 0.4em\relax Aegean Park Press, 1967.

\bibitem{gold1967optimal}
R.~Gold, ``{Optimal binary sequences for spread spectrum multiplexing
  (Corresp.)},'' \emph{IEEE Transactions on Information Theory}, vol.~13,
  no.~4, pp. 619--621, 1967.

\bibitem{kasami1966weight}
T.~Kasami, ``{Weight distribution formula for some class of cyclic codes},''
  \emph{Coordinated Science Laboratory Report no. R-285}, 1966.

\bibitem{rushanan2006weil}
J.~J. Rushanan, ``{Weil sequences: A family of binary sequences with good
  correlation properties},'' in \emph{2006 IEEE International Symposium on
  Information Theory}.\hskip 1em plus 0.5em minus 0.4em\relax IEEE, 2006, pp.
  1648--1652.

\bibitem{pott2006finite}
A.~Pott, \emph{{Finite geometry and character theory}}.\hskip 1em plus 0.5em
  minus 0.4em\relax Springer, 2006.

\bibitem{levanon2004radar}
N.~Levanon and E.~Mozeson, \emph{{Radar signals}}.\hskip 1em plus 0.5em minus
  0.4em\relax John Wiley \& Sons, 2004.

\bibitem{jedwab2006peak}
J.~Jedwab and K.~Yoshida, ``The peak sidelobe level of families of binary
  sequences,'' \emph{IEEE transactions on information theory}, vol.~52, no.~5,
  pp. 2247--2254, 2006.

\bibitem{nasrabadi2010survey}
M.~A. Nasrabadi and M.~H. Bastani, ``{A survey on the design of binary pulse
  compression codes with low autocorrelation},'' in \emph{Trends in
  Telecommunications Technologies}.\hskip 1em plus 0.5em minus 0.4em\relax
  IntechOpen, 2010.

\bibitem{lindner1975binary}
J.~Lindner, ``{Binary sequences up to length 40 with best possible
  autocorrelation function},'' \emph{Electronics letters}, vol.~11, no.~21, pp.
  507--507, 1975.

\bibitem{baden1990optimal}
J.~Baden and M.~Cohen, ``{Optimal peak sidelobe filters for biphase pulse
  compression},'' in \emph{IEEE International Conference on Radar}.\hskip 1em
  plus 0.5em minus 0.4em\relax IEEE, 1990, pp. 249--252.

\bibitem{coxson2005efficient}
G.~Coxson and J.~Russo, ``{Efficient exhaustive search for
  optimal-peak-sidelobe binary codes},'' \emph{IEEE Transactions on Aerospace
  and Electronic Systems}, vol.~41, no.~1, pp. 302--308, 2005.

\bibitem{leukhin2012binary}
A.~Leukhin and E.~Potehin, ``{Binary sequences with minimum peak sidelobe level
  up to length 68},'' \emph{arXiv preprint arXiv:1212.4930}, 2012.

\bibitem{leukhin2013optimal}
A.~N. Leukhin and E.~N. Potekhin, ``{Optimal peak sidelobe level sequences up
  to length 74},'' in \emph{2013 European Radar Conference}.\hskip 1em plus
  0.5em minus 0.4em\relax IEEE, 2013, pp. 495--498.

\bibitem{leukhin2014exhaustive}
{Leukhin, Anatolii N and Potekhin, Egor N}, ``{Exhaustive search for optimal
  minimum peak sidelobe binary sequences up to length 80},'' in
  \emph{International Conference on Sequences and Their Applications}.\hskip
  1em plus 0.5em minus 0.4em\relax Springer, 2014, pp. 157--169.

\bibitem{leukhin2015bernasconi}
A.~Leukhin and E.~Potekhin, ``{A Bernasconi model for constructing ground-state
  spin systems and optimal binary sequences},'' in \emph{Journal of Physics:
  Conference Series}, vol. 613, no.~1.\hskip 1em plus 0.5em minus 0.4em\relax
  IOP Publishing, 2015, p. 012006.

\bibitem{leukhin2017exhaustive}
A.~Leukhin, N.~Parsaev, V.~Bezrodnyi, and N.~Kokovihina, ``{The exhaustive
  search for optimum minimum peak sidelobe binary sequences},'' \emph{Bulletin
  of the Russian Academy of Sciences: Physics}, vol.~81, no.~5, pp. 575--578,
  2017.

\bibitem{NC}
C.~J. Nunn and G.~E. Coxson, ``{Best-known autocorrelation peak sidelobe levels
  for binary codes of length 71 to 105},'' \emph{IEEE transactions on Aerospace
  and Electronic Systems}, vol.~44, no.~1, pp. 392--395, 2008.

\bibitem{dzvonkovskaya2008long}
A.~Dzvonkovskaya and H.~Rohling, ``Long binary phase codes with good
  autocorrelation properties,'' in \emph{2008 International Radar
  Symposium}.\hskip 1em plus 0.5em minus 0.4em\relax IEEE, 2008, pp. 1--4.

\bibitem{Patent}
K.~L. Du, W.~H. Wu, and W.~H. Mow, ``{Determination of long binary sequences
  having low autocorrelation functions},'' Jul.~23 2013, uS Patent 8,493,245.

\bibitem{mow2015new}
W.~H. Mow, K.-L. Du, and W.~H. Wu, ``New evolutionary search for long low
  autocorrelation binary sequences,'' \emph{IEEE Transactions on aerospace and
  electronic systems}, vol.~51, no.~1, pp. 290--303, 2015.

\bibitem{dimitrov2020efficient}
M.~Dimitrov, T.~Baitcheva, and N.~Nikolov, ``Efficient generation of low
  autocorrelation binary sequences,'' \emph{IEEE Signal Processing Letters},
  vol.~27, pp. 341--345, 2020.

\bibitem{he2009designing}
H.~He, P.~Stoica, and J.~Li, ``Designing unimodular sequence sets with good
  correlations—including an application to mimo radar,'' \emph{IEEE
  Transactions on Signal Processing}, vol.~57, no.~11, pp. 4391--4405, 2009.

\bibitem{soltanalian2012computational}
M.~Soltanalian and P.~Stoica, ``Computational design of sequences with good
  correlation properties,'' \emph{IEEE Transactions on Signal processing},
  vol.~60, no.~5, pp. 2180--2193, 2012.

\bibitem{song2015sequence}
J.~Song, P.~Babu, and D.~P. Palomar, ``Sequence design to minimize the weighted
  integrated and peak sidelobe levels,'' \emph{IEEE Transactions on Signal
  Processing}, vol.~64, no.~8, pp. 2051--2064, 2015.

\bibitem{kerahroodi2017coordinate}
M.~A. Kerahroodi, A.~Aubry, A.~De~Maio, M.~M. Naghsh, and M.~Modarres-Hashemi,
  ``A coordinate-descent framework to design low psl/isl sequences,''
  \emph{IEEE Transactions on Signal Processing}, vol.~65, no.~22, pp.
  5942--5956, 2017.

\bibitem{turyn1968sequences}
R.~Turyn \emph{et~al.}, ``Sequences with small correlation,'' in \emph{Error
  correcting codes}.\hskip 1em plus 0.5em minus 0.4em\relax Wiley New York,
  1968, pp. 195--228.

\bibitem{wiener1964extrapolation}
N.~Wiener, \emph{Extrapolation, interpolation, and smoothing of stationary time
  series}.\hskip 1em plus 0.5em minus 0.4em\relax The MIT press, 1964.

\bibitem{lin2019efficient}
R.~Lin, M.~Soltanalian, B.~Tang, and J.~Li, ``Efficient design of binary
  sequences with low autocorrelation sidelobes,'' \emph{IEEE Transactions on
  Signal Processing}, vol.~67, no.~24, pp. 6397--6410, 2019.

\bibitem{dmitriev2007bounds}
D.~Dmitriev and J.~Jedwab, ``Bounds on the growth rate of the peak sidelobe
  level of binary sequences,'' \emph{Advances in Mathematics of
  Communications}, vol.~1, no.~4, p. 461, 2007.

\end{thebibliography}

\end{document}